\documentclass{appolb}
\usepackage{epsfig}

\begin{document}
\title{The phase diagram in the vector meson extended linear sigma model%
\thanks{CPOD 2016, Wroclaw}%
}
\author{Gy. Wolf and P. Kov\'acs
\address{Wigner Research Centre for Physics, Hungarian Academy of Sciences, 
H-1525 Budapest, POB 49, Hungary}
}
\maketitle
\begin{abstract}
  We investigate the chiral phase transition of the strongly interacting matter at nonzero temperature and baryon chemical potential $\mu_B$ within an extended (2+1) flavor Polyakov constituent quark-meson model which incorporates the effect of the vector and axial vector mesons. The parameters of the model are determined by comparing masses and tree-level decay widths with experimental values. We examine the restoration of the chiral symmetry by monitoring the temperature evolution of condensates. We study the T-$\rho_B$ phase diagram of the model and find that a critical end point exists, although at very low density.
\end{abstract}
  
\section{Introduction}
We investigate the thermodynamics of the strongly interacting matter
at high temperature and/or density with the ($2+1$) flavor Polyakov quark
meson model in which, beyond the vector and axial vector mesons
included alongside the scalar and pseudoscalar ones. We take into
account, as fermionic degrees of freedom, the constituent quarks
propagating on a constant gluon background in the temporal direction,
which naturally leads in a mean-field treatment to the appearance of
the Polyakov loop. We incorporate the vacuum fluctuations of the fermions in
the grand potential. We compare our results at $\mu_B=0$ with lattice data
and then move to finite $\mu_B$.
The details of the model can be found in \cite{elsm_2013,KSzW_2016}.

\section{Formulation of the Model}
\label{Sec:model}

We work with the Lagrangian used in \cite{KSzW_2016}. Compared to the usual
chiral Lagrangian, we introduce additional kinetic and Yukawa coupling terms
for the constituent quarks $\Psi = (q_u, q_d, q_s)^{T}$. Another important
modification is the presence of the gluon field in the covariant
derivative of the quark field. In the mean-field approximation, this
will give rise in the grand potential to
the appearance of the Polyakov loop, which mimics some properties of
the quark confinement. We use the logarithmic Polyakov loop potential
\cite{Fukushima:2003fw}. This parametrization does not include
the backreaction of the dynamical quarks on the gauge sector and therefore
the influence of the quarks on the deconfinement phase transition. In the
present study we shall use the improved Polyakov loop potential proposed in
\cite{Haas:2013qwp}.

There are altogether $16$ unknown parameters, $15$ parameters found in
the Lagrangian and the renormalization scale.
In the parametrization procedure we use alongside $29$ vacuum
quantities, and also the pseudocritical temperature $T_c$ at $\mu_B=0$ \cite{KSzW_2016}.
The value of the masses and decay constants are compared with the corresponding
experimental value taken from the PDG \cite{PDG}.
The $T_c$ pseudocritical temperature, which should be around
$150$~MeV, can reach very high values ($\geq 350$~MeV) in case of some
solutions.  Thus we chose to include the
physical value of $T_c$ in the parametrization with a $10\%$ error to throw
away solutions with high $T_c$, and unrealistic thermodynamics.

\subsection{The Grand Potential}
\label{sec:grand_pot}

To study the thermodynamics of a symmetric quark matter
($\mu_u=\mu_d=\mu_s\equiv\mu_q=\mu_B/3$), we shall use the grand
potential $\Omega(T,\mu_q)$ obtained from the partition function of a
three-dimensional spatially uniform system of volume $V$ in thermal
equilibrium at temperature $T=1/\beta.$ Since the Polyakov loop 
is treated at mean-field level, there is no integration over the
gluons and in this case the Polyakov loop potential is simply
added to the grand potential.

In the present case
the vacuum and thermal fluctuations for the fermions are taken into
account, while the mesonic vacuum fluctuations are neglected and the
effects of the lightest mesonic thermal fluctuations ($\pi$, $K$,
$f_0^L$) are included only in the pressure and the thermodynamical
quantities derived from it. In this approximation the grand potential reads
\begin{equation}
\Omega_\textnormal{H}(T,\mu_q) = U(\left<M\right>) + 
U(\langle\Phi\rangle,\langle\bar\Phi\rangle)
+ \Omega_{\bar q q}^{(0)}(T,\mu_q),
\label{Eq:grand_pot_H}
\end{equation}

where $U(\left<M\right>)$ is the tree-level meson potential,
$U(\langle\Phi\rangle,\langle\bar\Phi\rangle)$ is the Polyakov loop
potential and $\Omega_{\bar q q}^{(0)}$ is the contribution of the
fermions.

\subsection{The curvature meson masses}

The squared masses of the scalar and pseudoscalar mesons are calculated
from the
elements of the corresponding curvature matrix:
\begin{equation}
  m^2_{i,{ab}} = \frac{\partial^2 \Omega(T,\mu_q )}{\partial
    \varphi_{i,a} \partial \varphi_{i,b}}
  \bigg|_\textnormal{min}=\mathrm{m}^2_{i,ab}+\Delta m^2_{i,ab}+\delta
  m^2_{i,ab}, 
\label{Eq:M2i_ab}
\end{equation}
where the three terms on the right-hand side are as follows:
$\mathrm{m}^2_{i,ab}$ is the tree-level mass
matrix, and $\Delta
m^2_{i,ab}$ and $\delta m^2_{i,ab}$ are the contributions of the
fermionic vacuum and thermal fluctuations, respectively.

The field equations, which determine
the dependence on $T$ and $\mu_B=3\mu_q$ of the chiral condensates
$\phi_N$ and $\phi_S$ and the Polyakov loop variables $\Phi$ and
$\bar\Phi,$ are obtained by extremizing the grand potential,
\begin{equation} 
\frac{\partial\Omega_\textnormal{H}}{\partial \phi_N} =
\frac{\partial\Omega_\textnormal{H}}{\partial \phi_S} =
\frac{\partial\Omega_\textnormal{H}}{\partial \Phi} =
\frac{\partial\Omega_\textnormal{H}}{\partial \bar\Phi} = 0.
\end{equation}
To calculate various thermodynamical quantities we add the contribution of a
meson $b\in\{\pi, K, f_0^L\}$ to the pressure with the formula
\begin{equation}
\Delta p_b(T) = - n_{b}T\int\frac{d^3 p}{(2\pi)^3}\ln(1-e^{-\beta E_b(p)}), 
\end{equation}
where $E_b(p)=\sqrt{p^2 + m_b^2},$ with $m_b$ being the meson mass, and
$n_b$ is the meson multiplicity ($n_\pi=3,$ $n_K=4,$ and $n_{f_0^L}=1$).

\section{Results}
\label{Sec:result}

% thermal evolution of condensates and meson masses
\begin{figure*}[!t]
\begin{center}
\leavevmode
\epsfig{file=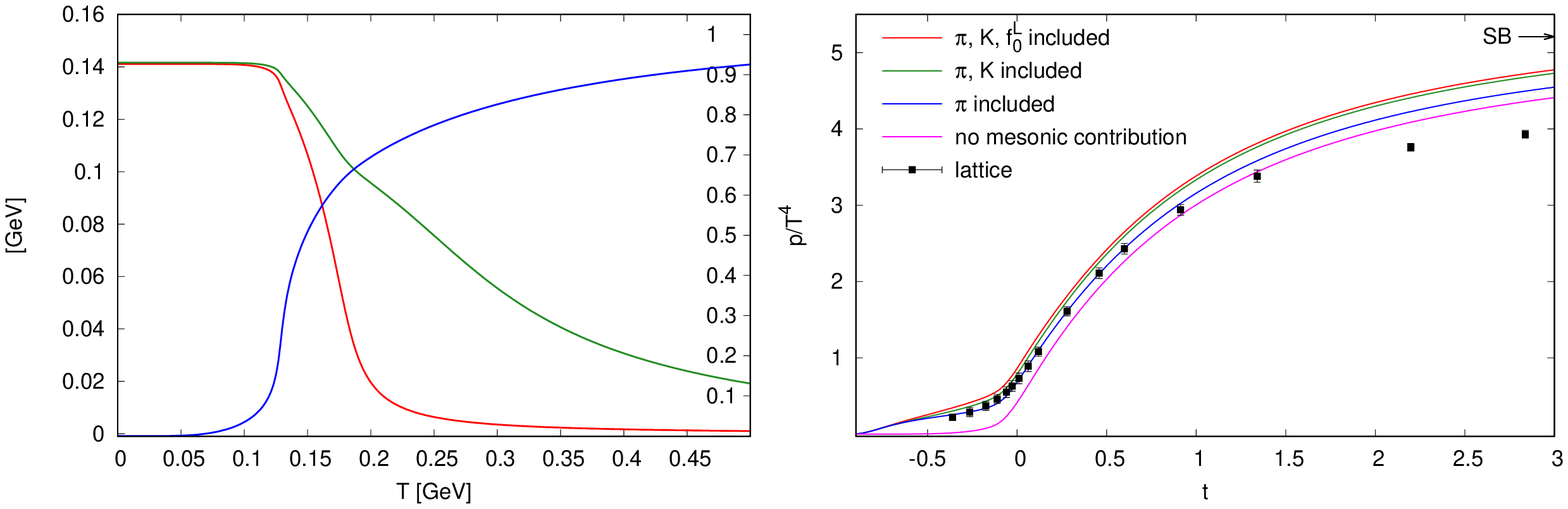,width=0.9\textwidth}
\caption{At $\mu_B=0$ temperature dependence of condensates from the improved
  Polyakov loop potential (left) and the pressure (right) as function of reduced temperature ($T/T_c-1$). Lattice data are taken from \cite{Borsanyi:2010cj}.
  Here $T_c^\mathrm{glue}=182$~MeV.}
\label{Fig:cond_mass}
\end{center}
\end{figure*}

On the left hand side of Fig.~\ref{Fig:cond_mass} we study at $\mu_B=0$ the temperature variation
of the nonstrange and strange chiral condensates and Polyakov loop expectation
value. We see that the chiral condensates stay close to their vacuum values up to some quite high value of the temperature of order 100~MeV.

On the right hand side of Fig.~\ref{Fig:cond_mass} we see that the
constituent quarks and the Polyakov loop potential give the major part of
the contribution to the pressure around and beyond $T_c$ and that at small
temperature the pressure is pion dominated. 
% INTERACTION MEASURE 

Some thermodynamical quantities derived from the pressure, like the
scale interaction measure $\Delta,$ the square of the speed of sound
$c_s^2$, and the equation of state parameter $p/\epsilon$ (pressure
over energy density), are shown in \cite{KSzW_2016} to be described very well by the model.
%%%
\subsection{$\mu_B-T$ phase diagram and the critical endpoint}
\label{SubSec:phase_diag}

We turn now to the study of the chiral phase transition at finite
quark chemical potentials: $\mu_u=\mu_d=\mu_s=\mu_q=1/3 \mu_B.$ The lattice
result of the curvature at $\mu_B=0$ is very well reproduced \cite{KSzW_2016}.
With the best set of parameters, the location of the CEP in our model is given
by $(\mu_B^\mathrm{CEP},T_c^\mathrm{CEP})=(885,52.7)$~MeV when the improved
Polyakov loop potential is used with $T^\mathrm{glue}_c=210$~MeV. The
phase diagram is similar to that obtained in \cite{Stiele:2016cfs},
with comparable values of the CEP's coordinates. The large
value of $\mu_B^\mathrm{CEP}$ we obtained is typical of a linear sigma
model without (axial)vector mesons in the case when the vacuum
fluctuation of the fermions is included. See, e.g.,
\cite{Chatterjee:2011jd} where the value $(\mu_B^\mathrm{CEP},T_c^\mathrm{CEP})=
(849,81)$~MeV was reported. Without the inclusion of the fermionic vacuum fluctuations, as
is the case of Refs.~\cite{Schaefer:2008hk,Mao:2009aq}, the value of
$\mu_B^{\mathrm{CEP}}$ is smaller and $T_c^{\mathrm{CEP}}$ higher,
compared to the case when they are properly taken into account.

\begin{figure}[!t]
\begin{center}
\leavevmode
\epsfig{file=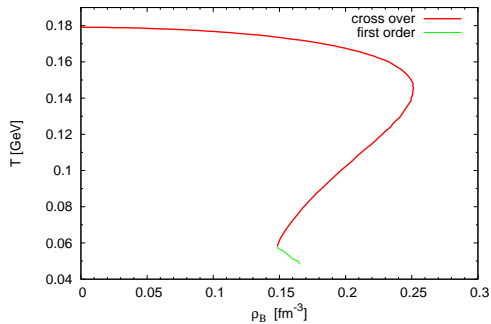,width=0.52\textwidth}
\caption{The phase diagram obtained by using the improved Polyakov
  loop potential. The dashed line shows the first order phase transition usinf Maxwell construction.}
\label{Fig:phase_bound}
\end{center}
\end{figure}
On Fig.~\ref{Fig:phase_bound} we show the phase diagram on the $T,\rho_B$ plane.
The dashed line shows the first order transition line whereas the solid curve is where the transition is cross over. The critical point is at rather low density, note, however, it is not the usual $\mu_u=\mu_d=\mu_q=1/3 \mu_B, \mu_s=0$ case.

\section{Conclusions}
\label{Sec:conclusion}

We have studied at finite temperature and baryonic densities the
thermodynamical properties of the Polyakov loop extended quark meson
model containing also vector and axial vector mesons. These latter
ingredients manifest themselves in a nontrivial way in the vacuum
parametrization of the model through their tree-level masses and decay
widths.  The $\chi^2$-minimization procedure was applied to fix the parameters.
We have studied the thermodynamics of the model, by using
an improved Polyakov loop potential, and obtained fairly good agreement with lattice data. We found that a CEP of the crossover transition line
exists in the $T-\rho_B$ phase diagram at rather low values of
$\rho_B.$

\section*{Acknowledgments}

The authors were supported by the Hungarian Research Fund (OTKA) under
Contract No.  K109462 and by the HIC for FAIR Guest Funds of the Goethe
University Frankfurt.

\end{document}